\RequirePackage{fix-cm}
\documentclass[sn-mathphys,Numbered]{sn-jnl}

\usepackage{amsbsy}
\usepackage{bm}
\usepackage{subcaption}
\pdfoutput=1
\usepackage{graphicx}%

\usepackage{multirow}%
\usepackage{tikz}
\usetikzlibrary{shapes,arrows}
\usetikzlibrary{intersections}
\usepackage{braket}
\usepackage{physics}
\usepackage{youngtab}
\usepackage{amsmath,amssymb,amsfonts}%
\usepackage{amsthm}%
\usepackage{mathtools}
\usepackage{bbm}
\usepackage{comment}
\usepackage{hyperref}
\usepackage{mathrsfs}%
\usepackage[T1]{fontenc} 
\usepackage[title]{appendix}%
\usepackage{xcolor}%
\usepackage{textcomp}%
\usepackage{manyfoot}%
\usepackage{booktabs}%
\usepackage{algorithm}%
\usepackage{algorithmicx}%
\usepackage{algpseudocode}%
\usepackage{listings}%

\def\q{{\mathbbm{q}}}
\newcommand{\Z}{{\mathbb Z}}

\def\ie{{\textit{i.e.}}}

\newcommand{\be}{\begin{equation}}
	\newcommand{\ee}{\end{equation}}

\theoremstyle{thmstyleone}%

\newtheorem*{proposition}{Proposition}%

\theoremstyle{thmstyletwo}%

\theoremstyle{thmstylethree}%
\newtheorem*{definition}{Definition}%
\newtheorem*{conjecture}{Conjecture}

\begin{document}

\title[Article Title]{\boldmath \textbf{Gukov-Pei-Putrov-Vafa conjecture for }$SU(N)/\mathbb{Z}_m$}

\author*{\fnm{Sachin} \sur{Chauhan}*}\email{sachinchauhan@iitb.ac.in}
\author{\fnm{Pichai} \sur{Ramadevi}}\email{ramadevi@iitb.ac.in}

\affil[]{\orgdiv{Department of Physics}, \orgname{Indian Institute of Technology}, \orgaddress{\street{Powai}, \city{Mumbai}, \postcode{400076}, \state{Maharashtra}, \country{India}}}

\abstract{In our earlier work, we studied the $\hat{Z}$-invariant(or homological blocks) for $SO(3)$ gauge group and we found it to be same as $\hat{Z}^{SU(2)}$. This motivated us to study the $\hat{Z}$-invariant for quotient groups $SU(N)/\mathbb{Z}_m$, where $m$ is some divisor of $N$. Interestingly, we find that $\hat{Z}$-invariant is independent of $m$.}


\maketitle

\section{Introduction}\label{sec:intro}

Over the past few decades, the notion of string dualities has emerged as a unifying thread connecting diverse branches of mathematics. These dualities, when expressed mathematically, have given rise to profound conjectures, fostering a deeper understanding of the connections between various mathematical domains. One notable duality in this context is the 3d-3d correspondence as documented in works by Dimofte and others\cite{Dimofte:2011ju,Dimofte:2011py,Dimofte:2010tz}. Such a correspondence establishes a link between complex Chern-Simons theory on a 3-manifold $M$ based on gauge group $G_{\mathbb{C}}$ and 3d $\mathcal{N}=2$ $T[M;G]$ theory.

This correspondence was a result of compactification of 6d (2,0) superconformal field theory(SCFT) of type ADE Lie algebra via topological twisting along 3-manifold $M$ so that $\mathcal{N}=2$ supersymmetry remains preserved in the remaining three flat directions:

\begin{equation}
	\text{6d}\; (2,0) \;\text{SCFT}\;\text{of type ADE Lie algebra}\;\stackrel{\text{compactification on M}}{\xrightarrow{\hspace{3.0cm}}} \;\text{3d} \;\mathcal{N}=2\;\; T[M;G].
\end{equation} 

\vspace{\baselineskip}
The data of 3d $\mathcal{N}=2$ $T[M;G]$ theory is given by manifold $M$. In other words, for every 3-manifold $M$ there would be a corresponding 3d $\mathcal{N}=2$ $T[M;G]$ theory encoding the geometry and topology of $M$. Many numerical and homological invariants of $M$ have been predicted by studying $T[M;G]$ on various backgrounds\cite{Gukov:2017kmk,Gukov:2016gkn}. Further, the topology and geometry of 3-manifolds is fairly well understood now\footnote{complete topological classification of 3-manifolds is still an open problem}, but still there is no known way to explicitly identify 3d $\mathcal{N}=2$ $T[M;G]$ for a general $M$.

In the seminal paper\cite{Witten:1988hf}, Witten showed that Chern-Simons theory is a topological gauge theory whose partition function $Z^G_{k'}[M;\q]$ contains the toplogical information of the underlying manifold $M$. Inspired by the work of Witten, Reshetikhin and Turaev constructed invariants of 3-manifolds using surgery prescription on links in $S^3$, commonly referred as Witten-Reshetikhin-Turaev(WRT) invariants\cite{Reshetikhin:1991tc}.

A concrete manifestation of the 3d-3d correspondence is the Gukov-Pei-Putrov-Vafa (GPPV) conjecture. This conjecture connects the partition function of Chern-Simons theory (or the Witten-Reshetikhin-Turaev (WRT) invariant) when evaluated on a 3-manifold $M$ to a specific type of invariant expressed in terms of $q$-series. In mathematical literature, these $q$-series valued invariants are commonly referred to as $\hat{Z}$ or homological blocks, and they are represented as vectors in the realm of $q$-series.

Initially, the exploration of $\hat{Z}$ was focused on $SU(2)$ gauge group through the analytic continuation of the WRT invariant defined for plumbed 3-manifold $M(\Gamma)$. This investigation led to the definition of $\hat{Z}$ for negative semidefinite plumbed 3-manifolds. Subsequent research efforts, as outlined in works\cite{2020higher,chauhan2022hat} extended the study of $\hat{Z}$ by examining the GPPV conjecture in the contexts of $SU(N)$, $SO(3)$, and $OSp(1|2)$ gauge groups.

Furthermore, it is essential to recognize that the WRT invariant for gauge group $G$, $\tau^{G}_{k'}[M(\Gamma);\q]$, is defined at a root of unity denoted as $\q\left(=\exp\left(\frac{2\pi i}{k'}\right)\right)$, where $k'$ signifies the renormalized Chern-Simons level. In contrast, the variable $q$ within $\hat{Z}^{G}_b[M(\Gamma);q]$ can be any complex number. Interestingly, as $q$ approaches $\q$, the relationship between these two invariants expressed through a $S$-transformation matrix\cite{Gukov:2016gkn,Gukov:2017kmk}:

\begin{equation}
	\tau^{G}_{k'}[M(\Gamma);\q]\cong \sum_{a,b}S_{ab}\hat{Z}^G_b[M(\Gamma);q]\Big\vert_{q\rightarrow\q},
\end{equation}

where $\hat{Z}_b^G[M(\Gamma);q]$ admits the following physical categorification:

\begin{equation}
	\hat{Z}_b^G[M(\Gamma);q]=\sum_{\substack{i\in \mathbb{Z}+\Delta_b\\j\in \mathbb{Z}}}q^i(-1)^j\text{dim}\;\mathcal{H}_{b,G}^{i,j}.
\end{equation}

In this equation, $\mathcal{H}_{b,G}^{i,j}$ corresponds to the BPS sector of the Hilbert space of $T[M;G]$ and $\Delta_b$ is a rational number specific to a 3-manifold $M$. These insights provide a promising avenue for addressing the long-standing categorification problem associated with the WRT invariant.

In our previous study\cite{chauhan2022hat}, the $\hat{Z}$-invariant for the $SO(3)$ gauge group was investigated by performing an analytical continuation of $\tau_{k}^{SO(3)}[M(\Gamma);\q]$ within a unit circle. Remarkably, it was discovered that $\hat{Z}^{SO(3)}_b[M(\Gamma);q]$ is equivalent to $\hat{Z}^{SU(2)}_b[M(\Gamma);q]$. This finding implies that the $\hat{Z}$-invariant depends on the Lie algebra rather than the Lie group, as $SU(2)$ and $SO(3)$ share the same Lie algebra. Furthermore, it is worth noting that $SO(3)\cong SU(2)/\mathbb{Z}_2$ is the Langlands dual group of $SU(2).$ Therefore, it raises the question of whether the equality $\hat{Z}^{SU(2)}_b[M(\Gamma);q] =\hat{Z}^{SO(3)}_b[M(\Gamma);q]$ can be attributed to this Langlands dual group correspondence. Additionally, for higher rank gauge group $SU(N)$ with $N>2$ and $N$ not being a prime number ($N\notin \mathbb{P}$), there exists gauge groups between $SU(N)$ and $SU(N)/\mathbb{Z}_N$. These groups are formed by taking a quotient of $SU(N)$ with a subgroup $\mathbb{Z}_m$ of $\mathbb{Z}_N$. All these quotient groups share the same $\mathfrak{su}(N)$ Lie algebra. Exploring the $\hat{Z}$-invariant for $SU(N)/\mathbb{Z}_m$ quotient groups will eventually answer whether $\hat{Z}$-invariant depends on $m$.

While the physics perspective suggests that $\hat{Z}$-invariant should be Lie algebra dependent only as 3d $\mathcal{N}=2$ $T[M;G]$ obtained by compactifying 6d $\mathcal{N}=(2,0)$ SCFT of type ADE Lie algebra on 3-manifold $M$. But in certain cases compactified theories does have Lie group dependence instead of Lie algebra\cite{freed2014relative}. In this paper, our primary objective is to address the question of whether the $\hat{Z}$-invariant exhibits dependence on Lie group or Lie algebra. To achieve this, we explicitly study the Gukov-Pei-Putrov-Vafa conjecture for gauge groups of the form $SU(N)/\mathbb{Z}_m$. For this, we must first define the WRT invariant for $SU(N)/\mathbb{Z}_m$ gauge group and then proceed with an analysis similar to that conducted in Ref\cite{2020higher}.

The organization of this paper is as follows: In section (\ref{review_3d3d}), we provide an overview of the GPPV conjecture and $\hat{Z}$-invariant. Section (\ref{wrtandlevel}) introduces the appropriate formula for the WRT invariant for the quotient group $SU(N)/\mathbb{Z}_m$. In section (\ref{refinement}), we demonstrate how to decompose the WRT invariant into $\hat Z$-invariant. Finally, we conclude in section (\ref{conclusion}) by discussing open problems and offering concluding remarks.

\newpage

\paragraph{Notations and conventions:} We use the following notation throughout this paper.\\

\begin{tabular}{r l}
	$M$:& 3-manifold\\
	$G$:& Gauge group\\
	$G_{\mathbb{C}}$:& Complex gauge group\\
	$\mathfrak{g}$:& Lie algebra\\
	$P$:& Weight lattice\\
	$P_+$:& Cone of dominant integer weights\\
	$\Lambda_{i}$:& Fundamental weight vector where $i=1,2,\ldots,r$\\
	$Q$:& Root lattice\\
	$P'$:& Intermediate lattice between root and weight lattice\\
	$(P')^\bullet$:& Dual of lattice $P'$\\
	$(\lambda,\mu)$:& Denotes the inner product between any two weight vectors, $\lambda$ and $\mu$\\
	$\Gamma$:& Plumbing graph of tree type\\
	$L$:& Number of vertices in a plumbing graph $\Gamma$\\
	$B$:& Linking matrix associated to plumbing graph $\Gamma$\\ 
	$b_{\pm}$:& Number of positive and negative eigenvalues of $B$\\
	$\sigma$:& Signature of linking matrix $B$ \ie~ $\sigma=b_+-b_-$\\
	$W$:& Weyl group\\
	$|W|$:& Order of the Weyl group\\
	$\omega_i$:& $i^{\text{th}}$ element of the Weyl group\\
	$\ell(\omega)$:& Length of Weyl group element $\omega$\\
	$\rho$:& Weyl vector\\
	$M(\Gamma)$:& Plumbed 3-manifold\\
	$k'$:& Renormalized Chern-Simons level\\
	$k$:& Bare Chern-Simons level\\
	$\q$:& Root of unity, $\exp(\frac{2\pi i}{k'})$\\
	$q$:& An arbitrary complex number inside the unit circle\\
	$\text{deg }v$:& Denotes the degree of vertex $v$ in a plumbing graph $\Gamma$\\ 
	$\tau^{G}_{k'}[M(\Gamma);\q]$:& WRT invariant for gauge group $G$\\ 
	$Z^G_{k'}[M;\q]$:& Chern-Simons partition function which is related to WRT invariant as $Z_{k'}^G[M;\q] = {\tau_{k'}^G[M;\q] \over \tau_{k'}^G[S^2 \times S^1;\q]}$\\
	$\hat{Z}^{G}_b[M(\Gamma);q]$:& $\hat{Z}$-invariant labelled by index $b$ for gauge group $G$
\end{tabular}

\section{Review of GPPV conjecture and $\hat{Z}$-invariant}
\label{review_3d3d}

In this section, we will give a brief survey of GPPV conjecture and $\hat{Z}$-invariant. Moreover, in this paper, we would limit ourselves to the case of rational homology sphere or plumbed 3-manifold coming from tree type diagrams only. Therefore, we first introduce the plumbed 3-manifolds:

\subsection{Plumbed 3-manifold}
\begin{figure}[h]
	\centering
	\includegraphics[height=1.2in,width=2.5in]{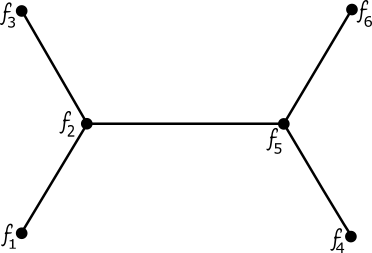}~~~~~~~~~
	\includegraphics[height=1.2in,width=2.5in]{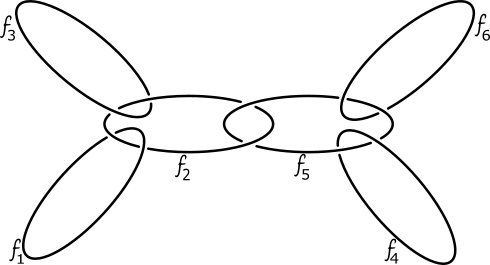}
	\caption{An example of a plumbing graph $\Gamma$ and the corresponding link $\mathcal{L}$}
	\label{Fig1}
\end{figure}

In general, any connected, closed, orientable 3-manifold can be obtained by surgery on a framed link in $S^3$\cite{10.2307/1970373,wallace_1960}. In this context, we focus our attention on the $L$-component link $\mathcal{L}$, composed of unknots with framing $f_i$. The resulting manifold, which emerges through a surgical operation on $\mathcal{L}$, is referred to as a plumbed 3-manifold. For example, we represent a six component link $\mathcal{L}$ through a graph $\Gamma$ with six vertices and call it a plumbing graph as shown in Figure \ref{Fig1}. 

The degree of any vertex $v$ ($\text{deg} \; v$) is equal to the total number of edges intersecting $v$.  In Figure \ref{Fig1}, we can see that vertices 2 and 5 have a degree of 3, while vertices 1, 3, 4, and 6 each have a degree of 1. Further, we associate a linking matrix $B$ with graph $\Gamma$. It is defined as follows:

\begin{equation}
	B_{v_1,v_2}=\begin{cases}
		1,~~~ v_1,v_2~ \text{connected},\\
		f_v,~~ v_1=v_2=v,\\
		0, ~~~\text{otherwise}
	\end{cases}
\end{equation}

Moreover, we define the signature of the linking matrix $B$ as $\sigma=b_+-b_-$ where $b_{\pm}$ denotes the number of positive and negative eigenvalues respectively. So, depending on the sign of $\sigma$ we referred the corresponding 3-manifold as positive or negative semidefinite plumbed 3-manifold $M$. Three-manifolds obtained from a plumbing graph $\Gamma$ or $\Gamma'$ using surgery prescription are homeomorphic if $\Gamma$ can be converted into $\Gamma'$ using the following set of transformations\cite{kirby1978calculus,neumann1981calculus}:
\begin{figure}[h]
	\centering
	\includegraphics[height=1.4in,width=6.2in]{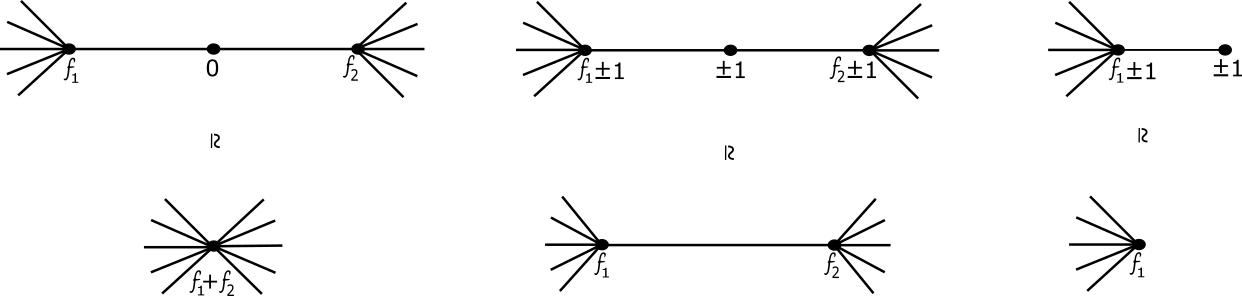}
	\caption{Kirby-Neumann moves which results in homeomorphic three-manifolds}
\end{figure}

\subsection{GPPV conjecture}

In Ref\cite{Gukov:2017kmk} Gukov et al. proposed a relation between WRT and $\hat{Z}$-invariant for negative-definite plumbed 3-manifold $M$. This relation was obtained using analytic continuation of $\tau_{k}^{SU(2)}[M(\Gamma),\q]$ and employing the Gauss sum reciprocity formula. The relation is as follows:

\begin{multline}
	\tau_{k}^{SU(2)}[M(\Gamma),\q]=\frac{1}{2\,(\q^{1/2}-\q^{-1/2})\,|\det B|^{1/2}}
	\,\times\\
	\sum_{a\in \mathrm{Coker}\,B}e^{-2\pi i(k+2)(a,B^{-1}a)}
	\sum_{b \in 2\mathrm{Coker}\, B+\delta} e^{-2\pi i(a,B^{-1}b)}\lim_{q\rightarrow \q\left(=\exp(\frac{2\pi i}{k+2})\right)} \hat{Z}_b^{SU(2)}[M(\Gamma);q],
	\label{WRT-prop-1}
\end{multline}

where $\delta=\{\delta_1,\delta_2,\ldots,\delta_L\}$ and $\delta_v = \text{deg }v \;(\text{mod 2})$. In literature, this relation is referred to as Gukov-Pei-Putrov-Vafa conjecture. Subsequently, this conjecture was studied for $SU(N)$,$SO(3)$ and $OSp(1|2)$ groups in Ref\cite{2020higher,chauhan2022hat,Costantino:2021yfd}. For the $SO(3)$ group, we deduced the conjecture in the following form:

\begin{multline}
	\tau_{k}^{SO(3)}[M(\Gamma);\q]=\frac{1}{2\,(\q^{1/2}-\q^{-1/2})\,|\det B|^{1/2}}
	\,
	\sum_{a\in \mathrm{Coker}\,B}e^{\textcolor{blue}{-\pi i(2k+1)}(a,B^{-1}a)}\\\\
	\sum_{b \in 2\mathrm{Coker}\, B+\delta} e^{-\pi i\big(a,B^{-1}(b{\color{blue}+BI})\big)}\lim_{q\rightarrow \q\left(=\exp(\frac{2\pi i}{4k+2})\right)} \hat{Z}^{SU(2)}_b[M(\Gamma);q]~.
	\label{gppvso3}
\end{multline}

From this expression, we see that $\hat{Z}$-invariant is same for both $SU(2)$ and $SO(3)$ groups. However, from equation (\ref{gppvso3}), we see that the overall factor undergoes modification, as highlighted in blue. Further, a proof of this conjecture for $SU(2)$ group appeared in Ref\cite{murakami2023proof}. We will now present a brief overview of the physical and mathematical definitions of the $\hat{Z}$-invariant.

\subsection{$\hat{Z}$-invariant}
$\hat{Z}$-invariants are $q$-series valued topological invariants of 3-manifold $M$. They admit the physical definition as a partition function of $T[M;SU(2)]$ evaluated on a cigar geometry $D^2\times_q S^1$ \ie 

\begin{equation}
	\hat{Z}^{SU(2)}_b[M;q]=Z_{T[M;SU(2)]}(D^2\times_q S^1;b)
\end{equation}

where $b$ is the index for these invariants and it belongs to $\text{Spin}^c(M)$\cite{gukov2021two}. This relation can also be written in the following way:

\begin{equation}
	\hat{Z}^{SU(2)}_b[M;q]=\sum_{\substack{i\in \mathbb{Z}+\Delta_b\\j\in \mathbb{Z}}}q^i(-1)^j\text{dim}\;\mathcal{H}_{b,SU(2)}^{i,j}
\end{equation}

where $\Delta_b$ is a rational number\footnote{usually referred to as delta invariant of $M$, see\cite{gukov2021cobordism,Harichurn:2023akp} for more} labeled by $\text{Spin}^c(M)$ and $\mathcal{H}_{b,SU(2)}^{i,j}$ corresponds to the BPS sector of the Hilbert space of $T[M;SU(2)]$. $\mathcal{H}_{b,SU(2)}^{i,j}$ are doubly-graded homological invariants of $M$. Hence the $\hat{Z}^{SU(2)}_b[M;q]$ are often referred to as homological blocks. Moreover one can take direct sum of all these homological invariants labelled by $b$ as follows:

\begin{equation}
	\mathcal{H}_{D^2,SU(2)}[M]=\bigoplus_{b\in \text{Spin}^c(M)/\mathbb{Z}_2}\;\;\bigoplus_{\substack{i\in\mathbb{Z}+\Delta_b,\\j\in\mathbb{Z}}}\mathcal{H}_{b,SU(2)}^{i,j}
\end{equation}

then this vector space $\mathcal{H}_{D^2,SU(2)}[M]$ is widely seen as the closed 3-manifold analog of Khovanov-Rozansky knot homology. From these definitions, we can interpret coefficients of powers of $q$ arising in $\hat{Z}$-invariant as counting the dimension of the Hilbert space of $T[M;SU(2)]$.

For negative semidefinite 3-manifold $M(\Gamma)$, $\hat{Z}$-invariant admits the following integral form for $\mathfrak{su}(2)$ Lie algebra\footnote{we are using the Lie algebraic notation because we know $\hat{Z}$ is same for $SU(2)$ and $SO(3)$ group}:

\begin{equation}
	\hat{Z}_b^{\mathfrak{su}(2)}[M(\Gamma);q]=(-1)^{b_+} q^{\frac{3\sigma-\sum_v f_{v}}{4}}\cdot\text{v.p.}\int\limits_{|z_v|=1}
	\prod_{v\;\in\; \text{Vertices}}
	\frac{dz_v}{2\pi iz_v}\,
	\left({z_v-1/z_v}\right)^{2-\text{deg}(v)}\cdot\left(\sum_{s \in 2B\Z^L+b}q^{-\frac{(s,B^{-1}s)}{4}}
	\prod_{i=1}^Lz_i^{s_i}\right),
\end{equation}

and ``v.p.'' means that we take principle value integral. Similar expression of $\hat{Z}$ for $SU(N)$, $OSp(1|2)$, $OSp(2|2)$ and $SU(2|1)$ for the plumbed 3-manifolds was found in Ref\cite{2020higher,chauhan2022hat,chae2021towards,Ferrari:2020avq}. Moreover, these invariants exhibits the quantum modularity\cite{Cheng:2023row,Cheng:2019uzc}. For $SU(N)$ group, $\hat{Z}$-invariant can be described as\cite{2020higher,Cheng:2018vpl}:

\begin{gather}
	\hat{Z}_b^{{\rm SU}(N)}[M(\Gamma);q] = (-1)^{\frac{N(N-1)}{2}b_+}q^{\frac{3\sigma -\operatorname{Tr}{\rm B}}{2}\frac{N^3-N}{12}}\\
	\hphantom{\hat{Z}_b^{{\rm SU}(N)}(Y;q) =}{} \times {\rm v.p.}\oint_{|z_{vj}|=1}\prod_{v\in V}\prod_{1\leq j\leq N-1}\frac{{\rm d}z_{vj}}{2\pi{\rm i} z_{vj}} F_{3d}(z)\Theta_{2d}^{b}(z,q)
	\label{sunzhat}
\end{gather}
with
\begin{gather*}
	F_{3d}(z) := \prod_{v\in V}\left(\sum_{\omega\in W}(-1)^{\ell(\omega)} \prod_{1\leq j\leq N-1}z_{vj}^{(\Lambda_{j},\omega(\rho))}\right)^{2-\deg v}\nonumber\\
	\hphantom{F_{3d}(z)}{} = \prod_{v\in V}\left( \prod_{1\leq j < k\leq N}\big(y_{vj}^{1/2}y_{vk}^{-1/2}-y_{vj}^{-1/2}y_{vk}^{1/2}\big) \right)^{2-\deg v},\\
	\Theta_{2d}^{b}(z,q) :=\sum_{s\in {\rm B} Q^{L}+b}q^{-\frac{1}{2}(s,{\rm B}^{-1}s)}\prod_{v\in V}\prod_{1\leq j\leq N-1}z_{vj}^{-(\Lambda_j,s_{v})},
\end{gather*}
where $z_{j} = \frac{y_{j}}{y_{j+1}}$. Here, the principal value integral ``v.p.'' implies the average over $|W|$ number of deformed contours, each associated with a Weyl chamber. In the following section, we will focus on the WRT invariant for quotient group $SU(N)/\mathbb{Z}_m$ which is necessary to study the corresponding $\hat{Z}$-invariant.

\section{WRT invariant for $SU(N)/\mathbb{Z}_m$}
\label{wrtandlevel}

For a 3-manifold $M(\Gamma)$, we define the WRT invariant\footnote{we use the normalization $\tau^G_{k'}[S^3,M]=1$} for quotient group $SU(N)/\mathbb{Z}_m$ as follows\cite{Dedushenko:2018bpp,Jeffrey:1992tk,Kaul:2000xe}:

\begin{equation}
	\tau^{SU(N)/\mathbb{Z}_m}_{k'}[M(\Gamma);\q]=\mathcal{\widetilde{S}}_{\rho\rho}^{L-1}\frac{\sum_{C^L}\prod_{v\in V}\mathcal V_v\prod_{e\in E}\mathcal E_e}{\left(\sum_{C}\mathcal V(+1\bullet)\right)^{b_+}\left(\sum_{C}\mathcal V(-1\bullet)\right)^{b_-}},
	\label{wrtA1}
\end{equation}

where $\mathcal{V}$, $\mathcal{E}$ denotes the vertex and edge factors of $L$-component plumbing graph $\Gamma$, $b_{\pm}$ represents the number of positive and negative eigenvalues of the linking matrix $B$ and $\pm1 \bullet$ denotes the single vertex with $\pm 1$ framing. The summation is performed over the set of allowed representations of the $SU(N)/\mathbb{Z}_m$ group, which are:

\begin{equation}
	C=\{\lambda\in (P_+\cap P')+\rho\;\vert\;(\lambda,\theta^\vee)<k'\}.
\end{equation}

Here, $P_+$ represents the set of dominant weights, $\theta^\vee$ refers to the maximal root, $\rho$ denotes the Weyl vector and $P'$ is a sublattice of $P$ such that there is an isomorphism between abelian group $P/P'$ and cyclic group $\mathbb{Z}_m(Q\subseteq P'\subseteq P)$. Furthermore, when $m=N$, then $P'$ is simply the root lattice $Q$, and when $m=1$, then $P'=P$. The vertex and edge factor can be expressed in terms of $\mathcal{\widetilde{S}}$ and $\mathcal{\widetilde{T}}$ matrices:

\begin{equation}
	\mathcal V_v=\mathcal{\widetilde{T}}_{\lambda\lambda}^{f_v}\mathcal{\widetilde{S}}_{\rho\lambda}^{2-\text{deg} v}\;,\;\;\mathcal E=\mathcal{\widetilde{S}}_{\lambda\mu}.
\end{equation}

The $\mathcal{\widetilde{S}}$ and $\mathcal{\widetilde{T}}$ matrices are:\footnote{these matrices are exactly the usual modular transformation matrices when $m=1$}

\begin{equation}
	\mathcal{\widetilde{S}}_{\lambda\mu}=\frac{i^{|\Delta_+|}}{|P'/k'Q|^{1/2}}\sum_{\omega\in W}(-1)^{\ell(\omega)}\q^{(\omega(\lambda),\mu)},\;\;\;\;\;\mathcal{\widetilde{T}}_{\lambda\mu}=\delta_{\lambda\mu}\q^{\frac{1}{2}(\lambda,\lambda)}\q^{-\frac{1}{2}(\rho,\rho)}
	\label{STmat}
\end{equation}

with
\begin{equation}
	\q=\exp\left(\frac{2\pi i}{k'}\right)~~~\text{and}~~~k'\in\mathbb{Z}^+,
\end{equation}

 $W$, $Q$ and $k'$ denotes the Weyl group, root lattice and renormalized Chern-Simons level respectively.

\paragraph{\underline{Chern-Simons level for $SU(N)/\mathbb{Z}_m$ WRT invariant}}
In Ref\cite{Dijkgraaf:1989pz}, it was shown that the three-dimensional Chern-Simons gauge theories with compact gauge group $G$ are classified by fourth cohomology group of the classifying space of the gauge group: $H^4(BG;\mathbb{Z})$. The classification parameter is the Chern-Simons level of the theory which is most commonly denoted by $k$. The level $k'$ is the renormalised Chern-Simons level which is related to the bare Chern-Simons level $k$ for $SU(N)$ gauge group as follows:

\begin{equation}
	k'=k+N.
\end{equation}

However for $SU(N)/\mathbb{Z}_m$ group which is non-simply connected, the relation between $k$ and $k'$ is as follows:

\begin{equation}
	k'=\gamma k+N,
\end{equation}

where $\gamma$ is some integer which can be calculated by considering the following short exact sequence:

\begin{equation}
	1\longrightarrow \mathbb{Z}_m\longrightarrow SU(N)\stackrel{\pi}{\longrightarrow} SU(N)/\mathbb{Z}_m\longrightarrow 1,
	\label{shortexseq}
\end{equation}

where $\mathbb{Z}_m$ is the subgroup of $\mathbb{Z}_N$. Let $\alpha$ and $\tilde{\alpha}$ be the generators of $H^4(B(SU(N));\mathbb{Z})$ and $H^4(B(SU(N)/\mathbb{Z}_m);\mathbb{Z})$ respectively. Then we have the following relation:

\begin{equation}
	B\pi^*(\tilde{\alpha})=\gamma\alpha,
\end{equation}

where $\pi^*$ is the pullback map of $\pi$ in equation (\ref{shortexseq}). The factor $\gamma$ is simply determined by comparing the images of $\tilde{\alpha}$ and $\alpha$ in the cohomology group $H^*(BT;\mathbb{Z})$ where $T$ is the maximal torus of rank $N-1$. So, the factor $\gamma$ is found to be the smallest integer for which the following equation is satisfied\cite{Dijkgraaf:1989pz}:

\begin{equation}
	\frac{\gamma}{2}(\Lambda_a,\Lambda_a)\in\mathbb{Z}, \;\;\forall \;a,
\end{equation}

where $\Lambda_a$'s are the fundamental weight vectors corresponding to the subgroup $\mathbb{Z}_m$ of $\mathbb{Z}_N$. For $SU(N)/\mathbb{Z}_N$ group, $\gamma$ is determined to be $2N$ when $N$ is even and $N$ when $N$ is odd:

\begin{equation}
	k' = \begin{cases}
		Nk+N,\;\;\;\;\;\;\text{when $N$ is odd}
		\\
		2Nk+N,\;\;\;\;\text{when $N$ is even}
	\end{cases}.
\end{equation}

For clarity, we have included the computation of sublattice $P'$ and Chern-Simons level $k'$ for certain non-simply connected groups in Appendix (\ref{appendB}). With this prescription of WRT for $SU(N)/\mathbb{Z}_m$, we will now focus on the corresponding $\hat{Z}$ by studying the GPPV conjecture.

\section{GPPV conjecture for $SU(N)/\mathbb{Z}_m$}
\label{refinement}

As discussed in the previous section, the WRT invariant $\tau_{k'}^{SU(N)/\mathbb{Z}_m}[M(\Gamma);\q]$ associated with $M(\Gamma)$ is given by

\begin{equation}
	\tau^{SU(N)/\mathbb{Z}_m}_{k'}[M(\Gamma);\q]=\mathcal{\widetilde{S}}_{\rho\rho}^{L-1}\frac{\sum_{C^L}\prod_{v\in V}\mathcal V_v\prod_{e\in E}\mathcal E_e}{\left(\sum_{C}\mathcal V(+1\bullet)\right)^{b_+}\left(\sum_{C}\mathcal V(-1\bullet)\right)^{b_-}}.
	\label{wrtA}
\end{equation}

For the sake of convenience, let's express the above equation in the following manner:

\begin{equation}
	\tau^{SU(N)/\mathbb{Z}_m}_{k'}[M(\Gamma);\q]=\frac{	\mathcal F[M(\Gamma);\q]}{\left(\mathcal F[M(+1\bullet);\q]\right)^{b_+}\left(\mathcal F[M(-1\bullet);\q]\right)^{b_+}},
\end{equation}

where

\begin{equation}
	\mathcal F[M(\Gamma);\q]=\frac{1}{(\mathcal{\widetilde{S}}_{\rho\rho})^{L+1}}\sum_{\substack{C^L}}\prod_{v\in V}\mathcal V_v\prod_{e\in E}\mathcal E_e.
	\label{F(Mgamma)}
\end{equation}

Similar to the $SU(2)$ group, we will have to perform Gauss decomposition of eqn.(\ref{wrtA}) to extract the homological blocks from it. Hence we rewrite the above equation (\ref{F(Mgamma)}) in a form so that we can use Gauss sum reciprocity formula\cite{DELOUP2007153}. We achieve this by extending the summation range $C^L$ over all Weyl chambers $W(C)^L$. Note that the matrices are invariant under the action of Weyl group elements.\footnote{upto a sign but that will not affect our final answer for $\tau^{SU(N)/\mathbb{Z}_m}_{k'}[M(\Gamma);\q]$} Hence we can sum over all the Weyl chambers and divide by the number of Weyl chambers to rewrite (\ref{F(Mgamma)}) as

\begin{equation}
	\mathcal F[M(\Gamma);\q]=\frac{1}{(\mathcal{\widetilde{S}}_{\rho\rho})^{L+1}|W|^L}\sum_{\substack{W(C)^L}}\prod_{v\in V}\mathcal V_v\prod_{e\in E}\mathcal E_e,~~~~~~~~~~~~~~~~~~~~~~~~~~~~~~~~~~~~~~~~~~~~~~~~~~~~~~~~~~~~~~~~~~
\end{equation}

\begin{multline}
	~~~~~~~~~~~~~~~~~~~~=\frac{1}{(\mathcal{\widetilde{S}}_{\rho\rho})^{L+1}|W|^L}\q^{-\frac{\sum_{j=1}^{L}f_j}{2}(\rho,\rho)}\left(\frac{i^{|\Delta_+|}}{|P'/k'Q|^{\frac{1}{2}}}\right)^{L+1}\sum_{\lambda\in W(C)^L}\underbrace{\prod_{v\in V}\left(\sum_{\omega\in W}(-1)^{\ell(\omega)}\q^{(\lambda_v,\omega(\rho))}\right)^{2-\text{deg}\;v}}_{\text{linear in }\lambda_v}\times\\\\
	\underbrace{\q^{\frac{f_v}{2}(\lambda_v,\lambda_v)}\prod_{(e_1,e_2)\in E}\sum_{\omega\in W}(-1)^{\ell(\omega)}\q^{(\omega(\lambda_{e_1}),\lambda_{e_2})}}_{\q^{\frac{1}{2}(\lambda,B\lambda)}}\;\;\;\;\;\;\;\;\;\;\;\;\;\;\;\;\;\;\;\;\;\;\;\;\;\;\;\;\;\;\;\;\;\;\;\;\;\;\;\;\;\;\;\;\;\;\;\;\;\;\;\;\;\;\;\;\;\;\;\;\;\;\;\;\;\;\;\;
	\label{wrtdec1}
\end{multline}

where we have used the fact that sum and product can be interchanged. The set over which the summation is being performed in equation (\ref{wrtdec1}) has now become $W(C)$. We further extend it to the whole lattice $((P\cap P')+\rho)/k'Q$ which is just $(P'+\rho)/k'Q$. In doing so, we observe for some representations $\lambda$, the term linear in $\lambda_v$  term (\ref{wrtdec1}) will be zero:

\begin{equation}
	\prod_{v\in V}\left(\sum_{\omega\in W}(-1)^{\ell(\omega)}\q^{(\lambda_v,\omega(\rho))}\right)^{2-\text{deg}\;v}=0.~~~~~~~~~~~~~~~~~~~~~~~~~~~~~~~~~~
	\label{linearterm}
\end{equation}

Using Weyl denominator formula, the expression can be rewritten as

\begin{equation}
	\prod_{v\in V}\left(\prod_{\alpha\in \Delta_+}\left(\q^{\frac{(\lambda_v,\alpha)}{2}}-\q^{-\frac{(\lambda_v,\alpha)}{2}}\right)\right)^{2-\text{deg}\;v}=0.~~~~~~~~~~~~~~~~~~~~~~~~~~~~\nonumber 
\end{equation}

Further, expressing $\lambda_v$ in terms of fundamental weight vectors $\Lambda_{i}$ \ie $\lambda_v=\sum_{j=1}^rn_{v_j}\Lambda_j$, the above equation becomes
	
\begin{equation}
~~~~~~~~\prod_{v\in V}\left(\prod_{\alpha\in\Delta_+}\left(\prod_{1\leq j\leq N-1}x_{v_j}^{\frac{(\Lambda_j,\alpha)}{2}}-\prod_{1\leq j\leq N-1}x_{v_j}^{-\frac{(\Lambda_j,\alpha)}{2}}\right)\right)^{2-\text{deg}\;v}\Bigg\vert_{x_{v_j}=\q^{n_{v_j}}}=0.
\end{equation}

 Hence, the points for which linear term in $\lambda_v$ becomes zero satisfy the following equation
 
 \begin{equation}
 	\sum_{1\leq j\leq N-1}n_{v_j}(\Lambda_j,\alpha)=0.
 \end{equation}
 
These points causes the singularity when $\text{deg }v>2$. Hence we first need to regularise the sum over these points. We introduce a parameter $\beta$ such that

\begin{equation}
	\beta\in\mathbb{C}~~~\text{and}~~~0< |\beta|< 1\nonumber.
\end{equation}

Using this parameter, we can rewrite the linear term in $\lambda_v$ as:

\begin{multline}
	~~~~~~~\prod_{v\in V}\left(\sum_{\omega\in W}(-1)^{\ell(\omega)}\q^{(\lambda_v,\omega(\rho))}\right)^{2-\text{deg}\;v}=
	\frac{1}{|W|^L}\prod_{v\in V}\Bigg[\left(\sum_{\omega\in W}(-1)^{\ell(\omega)}\q^{(\lambda_v,\omega(\rho))}\beta^{f(\omega,\omega_1)}\right)^{2-\text{deg}\;v}+\\\\ 
	\left(\sum_{\omega\in W}(-1)^{\ell(\omega)}\q^{(\lambda_v,\omega(\rho))}\beta^{f(\omega,\omega_2)}\right)^{2-\text{deg}\;v}+ \ldots+\left(\sum_{\omega\in W}(-1)^{\ell(\omega)}\q^{(\lambda_v,\omega(\rho))}\beta^{f(\omega,\omega_{|W|})}\right)^{2-\text{deg}\;v}\Bigg]\Bigg\vert_{\beta\rightarrow 1},
	\label{lineartermexpn}
\end{multline}

in which the function $f(\omega,\omega_i)$ is defined as follows:

\begin{equation}
	f(\omega,\omega_i):=\begin{cases}
		1, ~~~~~\text{when } \omega=\omega_i\\
		0, ~~~~~\text{otherwise}.
	\end{cases}
\end{equation}

The RHS of equation (\ref{lineartermexpn}) can be expanded as $|\beta|<1$:

\begin{equation}
	\frac{1}{|W|^L}\sum_{m\ge 0}\left(\sum_{s}\chi_s^m\q^{(\lambda,s)}\right)\beta^m,	
	\label{reg1}
\end{equation}

where $s=\{s_1,s_2,\ldots,s_L\}$ is some subset of $P^L$. Further, we interchange the summation to rewrite equation (\ref{reg1}) as

\begin{equation}
	\frac{1}{|W|^L}\sum_{s\in Q^L+\delta}\underbrace{\left(\sum_{m\ge 0}\chi_s^m\beta^m\right)}_{\xi^\beta_s}\q^{(\lambda,s)}=\frac{1}{|W|^L}\sum_{s\in Q^L+\delta}\xi_s^\beta \q^{(\lambda,s)},
\end{equation}

where $\chi_s^m\in \mathbb{Z}$. Hence, we can rewrite the linear term as series in $\q$, and  its coefficients $\xi_s^1$ can be determined by the following equation:

\begin{equation}
	\prod_{v\in V}\left(\sum_{\omega\in W}(-1)^{\ell(\omega)}\q^{(\lambda_v,\omega(\rho))}\right)^{2-\text{deg}\;v}=\frac{1}{|W|^L}\sum_{s\in Q^L+\delta}\xi_{s}^\beta\q^{(\lambda,s)}\Big\vert_{\beta\longrightarrow 1},
	\label{regularisation}
\end{equation}

where $\delta_v= (2-\text{deg}~v)\rho~ \text{mod}~Q$ and $\xi_s^1\in\mathbb{Z}$. In summary, we have rewritten the linear term as some series in $\q$. This series is obtained by taking an average of individual geometric series in $\q$, each determined by a specific selection of Weyl chamber. This completes our regularization of linear term. For clarity we have provided a detailed example in the appendix (\ref{appenda}). This led us to the following equation (\ref{beforereciprocity1}):

\begin{multline}
	~~~~~~~~~\mathcal F[M(\Gamma);\q]=\frac{1}{(\mathcal{\widetilde{S}}_{\rho\rho})^{L+1}|W|}q^{-\frac{\sum f_j}{2}(\rho,\rho)}\left(\frac{i^{|\Delta_+|}}{|P'/k'Q|^{\frac{1}{2}}}\right)^{L+1}\times\\\\
	\sum_{\lambda\in (P'+\rho)^{L}/k'Q^L}\left(\frac{1}{|W|^L}\sum_{s\in Q^L+\delta}\xi_{s}^\beta\q^{(\lambda,s)}\right)\q^{\frac{1}{2}(\lambda,B\lambda)}\Big\vert_{\beta\longrightarrow 1}.~~~~~~~~
	\label{beforereciprocity1}
\end{multline}

Now, in order to use the reciprocity formula we replace $Q$ with $\eta P'$ for some positive integer $\eta$ as $\eta P'\subseteq Q\subseteq P'\subseteq P$ and subsequently multiply it by the suitable factor given by the order of quotient of these two lattices. Hence, the above equation becomes:

\begin{multline}
	~~~~~~~~~\mathcal F[M(\Gamma);\q]=\frac{1}{(\mathcal{\widetilde{S}}_{\rho\rho})^{L+1}|W|}q^{-\frac{\sum f_j}{2}(\rho,\rho)}\left(\frac{i^{|\Delta_+|}}{|P'/k'Q|^{\frac{1}{2}}}\right)^{L+1}\underbrace{\left(\frac{1}{|Q/\eta P'|^{L}}\right)}_{\text{factor which compensates for replacing}~Q~\text{with}~\eta P'}\times\\\\
	\sum_{\lambda\in (P'+\rho)^{L}/\eta k'(P')^L}\left(\frac{1}{|W|^L}\sum_{s\in Q^L+\delta}\xi_{s}^\beta\q^{(\lambda,s)}\right)\q^{\frac{1}{2}(\lambda,B\lambda)}\Big\vert_{\beta\longrightarrow 1}.~~~~~~~~
	\label{beforereciprocity2}
\end{multline}

Since we are interested in non-simply connected group $SU(N)/\mathbb{Z}_m$ for which $\rho\notin P'$, we have to do the following shift in $\lambda$: $\lambda\longrightarrow \lambda+\boldsymbol{\rho}$, where $\boldsymbol{\rho}=(\underbrace{\rho,\rho,\ldots,\rho}_{L\text{-times}})$. Subsequently, we get the following:

\begin{multline}
	~~~~~~~~\mathcal F[M(\Gamma);\q]=\frac{1}{(\mathcal{\widetilde{S}}_{\rho\rho})^{L+1}|W|^{L+1}}\q^{-\frac{\sum f_j}{2}(\rho,\rho)}\left(\frac{i^{|\Delta_+|}}{|P'/k'Q|^{\frac{1}{2}}}\right)^{L+1}\left(\frac{1}{|Q/\eta P'|^{L}}\right)\q^{\frac{1}{2}(\boldsymbol\rho,B\boldsymbol\rho)}\times\\\\
	\sum_{s\in Q^L+\delta}\xi_s^\beta \q^{(\boldsymbol\rho,s)}\sum_{\lambda\in (P')^L/\eta k'(P')^L}\q^{\frac{1}{2}(\lambda,B\lambda)}\q^{(\lambda,s+B\boldsymbol\rho)}\Big\vert_{\beta\longrightarrow 1}.~~~~~~~~~~~
\end{multline}
\\
Now using Gauss sum reciprocity formula\cite{DELOUP2007153} and with the assumption that the quadratic form, $B:\mathbb{Z}^L\times \mathbb{Z}^L\longrightarrow \mathbb{Z}$ is negative definite\footnote{that is $\sigma=-L$},\footnote{in following equation $l$ denotes the rank of the lattice $(P')^L$ and $\ell$ represents the length of the Weyl group element}, $\mathcal F[M(\Gamma);\q]$ equals:

\begin{multline}
	=\left(\frac{1}{|W|\sum_{\omega\in W}(-1)^{\ell(\omega)}\q^{(\rho,\omega(\rho))}}\right)^{L+1}\q^{-\frac{\sum f_j}{2}(\rho,\rho)}\left(\frac{1}{|Q/\eta P'|^L}\right)\left(\frac{\exp(\frac{\pi i\sigma}{4})\;(\eta k')^{l/2}}{|\text{det}(\eta B)|^{\frac{N-1}{2}}\;\;\text{Vol}[((P')^\bullet)^L]}\right)\times\\\\
	\sum_{a\in ((P')^\bullet)^L/\eta B((P')^\bullet)^L}\exp[-\pi ik'(a,B^{-1}a)]\sum_{b\in (Q^L+\delta)/BQ^L}\exp[-2\pi i(a,B^{-1}(b+B\boldsymbol\rho))]\sum_{s\in BQ^L+b}\xi_s^\beta \q^{-\frac{(s,B^{-1}s)}{2}}\Big\vert_{\beta\longrightarrow 1},
\end{multline}

where $(P')^\bullet$ denotes the dual lattice of $P'$. The WRT invariant $\tau^{SU(N)/\mathbb{Z}_m}_{k'}[M(\Gamma);\q]$ including the framing factor reduces to

\begin{multline}
	=\frac{1}{|W|^{L+1}}\q^{-\frac{(\rho,\rho)}{2}(3L+\text{Tr} B)}\underbrace{\left(\sum_{a\in (P')^\bullet/\eta(P')^\bullet}\exp(\pi ik'(a,a)-2\pi i(a,\rho))\right)^{-L}}_{|(P')^\bullet/\eta(P')^\bullet|^{-L}}
	\frac{1}{|\text{det} B|^{\frac{N-1}{2}}\sum_{\omega\in W}(-1)^{\ell(\omega)}\q^{(\rho,\omega(\rho))}}\times\\\\
	\underbrace{\sum_{a\in ((P')^\bullet)^L/\eta B((P')^\bullet)^L}}_{|(P')^\bullet/\eta(P')^\bullet|^L\sum_{a\in ((P')^\bullet)^L/B((P')^\bullet)^L}}\exp(-\pi ik'(a,B^{-1}a))\sum_{b\in (Q^L+\delta)/BQ^L}\exp(-2\pi i(a,B^{-1}(b+B\rho)))\times\\\\
	\sum_{s \in BQ^L+b}\xi_s^\beta\q^{-\frac{(s,B^{-1}s)}{2}}\Big\vert_{\beta\longrightarrow 1},~~~~~~~~~~~~~~~~~~~~~~~~~~~~~~~~~~~~~~~~~~~~~~~~~~~~~~~~~~~~~~~~~~~~~~~~~~~~
\end{multline}

which simplifies to the following:

\begin{multline}
		\tau^{SU(N)/\mathbb{Z}_m}_{k'}[M(\Gamma);\q]=\frac{1}{|W|^{L+1}}\;
		\frac{q^{-\frac{(\rho,\rho)}{2}(3L+\text{Tr} B)}}{|\text{det} B|^{\frac{N-1}{2}}\sum_{\omega\in W}(-1)^{\ell(\omega)}q^{(\rho,\omega(\rho))}}\sum_{a\in ((P')^\bullet)^L/B((P')^\bullet)^L}\exp(-\pi ik'(a,B^{-1}a))\times\\\\
		\sum_{b\in (Q^L+\delta)/BQ^L}\exp(-2\pi i(a,B^{-1}(b+B\boldsymbol\rho)))\sum_{s \in BQ^L+b}\xi_s^\beta \q^{-\frac{(s,B^{-1}s)}{2}}\Big\vert_{\beta\longrightarrow 1}.~~~~~~~~~~~~~
\end{multline}

Now, assuming that the following holds:

\begin{equation}
	\lim_{\beta\longrightarrow 1}\sum_{s \in BQ^L+b}\xi_s^\beta \q^{-\frac{(s,B^{-1}s)}{2}}=\lim_{q\rightarrow \q}\sum_{s \in BQ^L+b}\xi_s^1 q^{-\frac{(s,B^{-1}s)}{2}},
\end{equation}

we finally obtain,

\begin{multline}
	\tau^{SU(N)/\mathbb{Z}_m}_{k'}[M(\Gamma);\q]=\frac{1}{|W|^{L+1}}\;
	\frac{q^{-\frac{(\rho,\rho)}{2}(3L+\text{Tr} B)}}{|\text{det} B|^{\frac{N-1}{2}}\sum_{\omega\in W}(-1)^{\ell(\omega)}q^{(\rho,\omega(\rho))}}\sum_{a\in ((P')^\bullet)^L/B((P')^\bullet)^L}\exp(-\pi ik'(a,B^{-1}a))\times\\\\
	\sum_{b\in (Q^L+\delta)/BQ^L}\exp(-2\pi i(a,B^{-1}(b+B\boldsymbol\rho)))\sum_{s \in BQ^L+b}\xi_s^1 q^{-\frac{(s,B^{-1}s)}{2}}\Big\vert_{q\longrightarrow \q},~~~~~~~~~~~~~
	\label{gaussdecomposed}
\end{multline}

\begin{multline}
	~~~~~~~~~~~~~~~~~~~~~~~~~=\frac{1}{|W||\text{det} B|^{\frac{N-1}{2}}\sum_{\omega\in W}(-1)^{\ell(\omega)}\q^{(\rho,\omega(\rho))}}\sum_{a\in ((P')^\bullet)^L/B((P')^\bullet)^L}\exp(-\pi
	ik'(a,B^{-1}a))\times\\\\
	\sum_{b\in (Q^L+\delta)/BQ^L}\exp(-2\pi i(a,B^{-1}(b+B\boldsymbol\rho)))\lim_{q\rightarrow \q}	\underbrace{\hat{Z}^{\mathfrak{su}(N)}_b[M(\Gamma);q]}_{\text{independent of }m}.~~~~~~~~~~~~~~~~~~~~~
\end{multline}

From equation (\ref{gaussdecomposed}) explicit expression of $\hat{Z}$-invariant can be read off as:

\begin{equation}
	\hat{Z}^{\mathfrak{su}(N)}_b[M(\Gamma);q]=|W|^{-L}q^{-\frac{(3L+\text{Tr}\;B)}{2}(\rho,\rho)}\sum_{s\in BQ^L+b}\xi_s^1 q^{-\frac{(s,B^{-1}s)}{2}}\;\;\in\;|W|^{-L}q^{\Delta_b}\mathbb Z[[q]],
\end{equation}

Thus we have shown that $\hat{Z}$-invariant does not depend on $m$. The overall factor which relates the $\hat Z$ with $\tau^{SU(N)/\mathbb{Z}_m}_{k'}[M(\Gamma);\q]$ has the $m$ dependence. This led us to the following proposition:\\

\begin{proposition}
	Let $M(\Gamma)$ be a negative definite plumbed 3-manifold. For non-simply connected group $SU(N)/\mathbb{Z}_m$, WRT invariant can be decomposed in the following form:
	\begin{multline}
	\tau^{SU(N)/\mathbb{Z}_m}_{k'}[M(\Gamma);\q]=\frac{1}{|W||\text{det} B|^{\frac{N-1}{2}}\sum_{\omega\in W}(-1)^{\ell(\omega)}\q^{(\rho,\omega(\rho))}}\sum_{a\in ((P')^\bullet)^L/B((P')^\bullet)^L}\exp(-\pi
	ik'(a,B^{-1}a))\times~~~~\\\\
\sum_{b\in (Q^L+\delta)/BQ^L}\exp(-2\pi i(a,B^{-1}(b\textcolor{blue}{+B\boldsymbol\rho})))\lim_{q\rightarrow \q}	\hat{Z}^{\mathfrak{su}(N)}_b[M(\Gamma);q],~~~~~~~~~~~
\label{proposition}
	\end{multline}

where $\bullet$ denotes the dual operation on the lattice $P'$, $k'=\gamma k+N$, and $\q=\exp\left(\frac{2\pi i}{k'}\right)$.\\
\end{proposition}

 Moreover, we can express the terms appearing as coefficients to $\hat{Z}$-invariant as linking pairing and homology group. The linking pairing is defined as follows:\\

\begin{definition}[Linking pairing]
For a closed and connected 3-manifold $M$, with $\partial M=\emptyset$, we have the linking pairing($\ell k$) on the torsion part of $H_1(M;\mathbb{Z})$,

\begin{equation}
	\ell k: \text{Tor}~H_1(M;\mathbb{Z})\otimes \text{Tor}~H_1(M;\mathbb{Z})\longrightarrow \mathbb{Q}/\mathbb{Z}.
\end{equation}

For $a,b\in \text{Tor}~H_1(M;\mathbb{Z})$, $\ell k$ is given as:

\begin{equation}
\ell k(a,b)=\frac{\#(a'\cdot b)}{n}~\text{mod}~\mathbb{Z},
\end{equation}

where $n\in \mathbb{Z}_{\neq 0}$ such that $n~a=0\in H_1(M;\mathbb{Z})$ and $a'$ is a 2-chain which is bounded as $\partial a'=na$. For plumbed 3-manifold $M(\Gamma)$, $\ell k$ is simply,

\begin{equation}
	\ell k(a,b)=(a,B^{-1}b)~\text{mod}~\mathbb{Z},~~~~a,b\in \mathbb{Z}^L/B\mathbb{Z}^L.
\end{equation}

\end{definition}

\vspace{0.5cm}
Using this we write the Gukov-Pei-Putrov-Vafa conjecture for $SU(N)/\mathbb{Z}_m$ as follows:\\

\begin{conjecture}
Let $M$ be a closed 3-manifold with $b_1(M)=0$ and $\text{Spin}^c(M)$ be the set of $\text{Spin}^c$ structures on $M$. Then WRT invariant $\tau^{SU(N)/\mathbb{Z}_m}_{k'}[M;\q]$ can be decomposed as follows:

\begin{align}
\tau^{SU(N)/\mathbb{Z}_m}_{k'}[M;\q]=\frac{1}{|H_1(M;\mathbb{Z})|^{\frac{N-1}{2}}\sum_{w\in W}(-1)^{\ell(w)}\q^{(\rho,\omega(\rho))}}\sum_{a,b\in (\text{Spin}^c(M))^{(N-1)}/S_N}\exp(-2\pi i k'\sum_{i=1}^{N-1}\ell k(a_i,a_i))\times\nonumber\\\nonumber\\
\textcolor{blue}{\exp(-2\pi i\sum_{i=1}^{N-1}a_i)}\exp(-4\pi i\sum_{i=1}^{N-1}\ell k(a_i,b_i))\lim_{q\rightarrow\q}\hat{Z}_b^{\mathfrak{su}(N)}[M;q],
\end{align}

where $\hat{Z}_b^{\mathfrak{su}(N)}[M;q]\in |W|^{-c}q^{\Delta_b}\mathbb{Z}[[q]]$ and $S_N$ is the symmetric group of degree $N$.\\
\end{conjecture}

Note that for simply connected $SU(N)$ group, appendix(\ref{appenc}), there is no shift in $b$. The shift in eqn.(\ref{proposition}), $b\longrightarrow b+B\boldsymbol\rho$, is attributed to the non-simply connected nature of $SU(N)/\mathbb{Z}_m$ group. This introduces the term $\exp(-2\pi i\sum_{i=1}^{N-1}a_i)$ in the above conjecture.


\section{Conclusions and future directions}
\label{conclusion}
In this paper, we have worked out the explict form of GPPV conjecture for the case of $SU(N)/\mathbb{Z}_m$ gauge group. We have found that the $\hat{Z}$-invariant is independent of $\mathbb{Z}_m$ factor. In fact, it turns out that the dependence of $\mathbb{Z}_m$ arises as an overall factor to WRT invariant:

\begin{equation}
	\tau^{SU(N)/\mathbb{Z}_m}_{k'}[M;\q]=\sum_{b}c_b(\mathbb{Z}_m)\hat{Z}_b^{\mathfrak{su}(N)}[M;q]\Big|_{q\rightarrow e^{\frac{2\pi i}{k'}}}.
\end{equation}

We list some of the issues which we encountered while doing this exercise. We hope to resolve these in our future works:

\begin{itemize}
	\item In the process of Gauss decomposition of WRT invariant, there exists singularities correspondingly to the walls of the Weyl group. For certain cases of quotient groups $SU(N)/\mathbb{Z}_m$, these singularities do not arise by definition(For eg. $SU(2)/\mathbb{Z}_2$). We are interested in observing the progression of the proof for the GPPV conjecture in these particular instances. Although, recently a proof of this conjecture appeared for simply laced Lie alegbras\cite{murakami2023homological} but the proof is not available for non-simply connected groups or quotient groups.\\
	
	\item We have conjectured the relation between WRT invariant for $SU(N)/\mathbb{Z}_m$ group and $\hat{Z}^{\mathfrak{su}(N)}$. It would be interesting to study the $w$-refined version of $SU(N)$ WRT invariant and its relation with $\hat{Z}^{\mathfrak{su}(N)}$ invariant. For $SU(2)$ WRT invariant, a $w$-refined WRT invariant was introduced in Ref.\cite{10.2140/agt.2015.15.1363}, and its relation with $\hat{Z}$ was studied in Ref.\cite{Costantino:2021yfd}.\\

	
	\item It would be nice to explore the extension of our work to more general 3-manifolds and knot complements.
\end{itemize}

\backmatter

\bmhead{Acknowledgments}

SC expresses gratitude to Pavel Putrov and Sunghyuk Park for numerous fruitful discussions. SC is also appreciative of the MoU involving CAS-IITB and ICTP, which provided the opportunity for a two-month visit to ICTP, where significant progress related to this research was achieved. SC extends thanks to the organizers of String-Math 2023, where a portion of this work was presented. Additionally, SC would like to acknowledge the IoE cell at IIT Bombay for providing financial support during the visit to String-Math 2023. SC and PR would like to thank all the speakers as well as the organisers of the Learning workshop on BPS states and 3-manifolds for discussions and interactions. PR would like to acknowledge the ICTP’s Associate programme where we made some progress during her visit as senior associate.

\begin{appendices}

\section{Explicit calculation of $\hat{Z}^{\mathfrak{su}(3)}$-invariant for a particular plumbing graph}
\label{appenda} 

In this appendix, we work out explicitly the $\hat{Z}$ for $\mathfrak{su}(3)$ Lie algebra for the following plumbed 3-manifold:

\begin{figure}[h]
	\centering
	\includegraphics[scale=0.5]{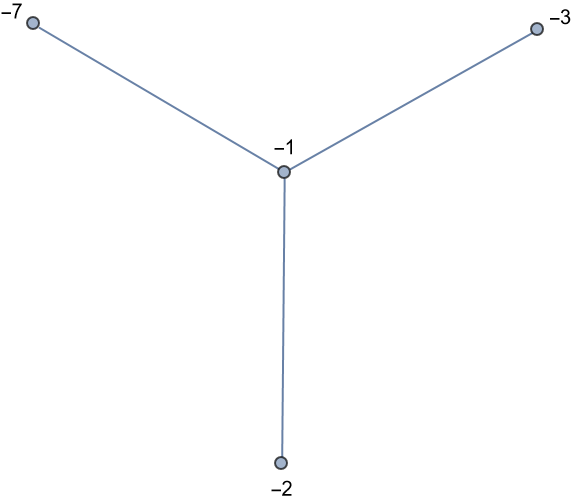}
\end{figure}

For $\mathfrak{su}(3)$ case, $W=\{1,s_1,s_2,s_1s_2,s_2s_1,s_1s_2s_1\}$, $\rho=\Lambda_1+\Lambda_2$ and $(\rho,\rho)=2$.

So, the $\hat{Z}$-invariant is:
\begin{equation}
	\hat{Z}^{\mathfrak{g}}_b[M(\Gamma),q]=|W|^{-L}q^{-\frac{(3L+\text{Tr}\;B)}{2}(\rho,\rho)}\sum_{s\in BQ^L+b}\xi_s^1 q^{-\frac{(s,B^{-1}s)}{2}}
	\label{zhatdef}
\end{equation} 

which in this case:

\begin{equation}
	\hat{Z}^{\mathfrak{su}(3)}_b[M(\Gamma),q]= \frac{q}{6^4}\sum_{s\in BQ^L+b}\xi_s^1 q^{-\frac{(s,B^{-1}s)}{2}}
\end{equation} 

where $\xi_s^1$ is determined from the following equation:

\begin{equation}
	\prod_{v\in V}\left(\sum_{\omega\in W}(-1)^{\ell(\omega)}\q^{(\lambda_v,\omega(\rho))}\right)^{2-\text{deg}\;v}=\frac{1}{6^4}\sum_{s\in Q^L+\delta}\xi_{s}^\beta\q^{(\lambda,s)}\Big\vert_{\beta\longrightarrow 1}
	\label{regularisation}
\end{equation}

We write the LHS of above equation as follows:

\begin{align}
\prod_{v\in V}\Big(\q^{(n^v_1\Lambda_1+n^v_2\Lambda_2)(\Lambda_1+\Lambda_2)}-\q^{(n^v_1\Lambda_1+n^v_2\Lambda_2)(-\Lambda_1+2\Lambda_2)}
-\q^{(n^v_1\Lambda_1+n^v_2\Lambda_2)(2\Lambda_1-\Lambda_2)}\nonumber\\+\q^{(n^v_1\Lambda_1+n^v_2\Lambda_2)(-2\Lambda_1+\Lambda_2)}
+\q^{(n^v_1\Lambda_1+n^v_2\Lambda_2)(\Lambda_1-2\Lambda_2)}-\q^{(n^v_1\Lambda_1+n^v_2\Lambda_2)(-\Lambda_1-\Lambda_2)}\Big)^{2-\text{deg }v}
\end{align}

where $\lambda_v=n^v_1\Lambda_1+n^v_2\Lambda_2$. The above equation simplies to the following using $\Lambda_1^2=\Lambda_2^2=\frac{2}{3}$ and $\Lambda_1\Lambda_2=\frac{1}{3}$:

\begin{align}
	\prod_{v\in V}\Big(\q^{(n^v_1+n^v_2)}-\q^{n^v_2}-\q^{n^v_1}+\q^{-n^v_1}+\q^{-n^v_2}-\q^{-(n^v_1+n^v_2)}\Big)^{2-\text{deg }v}
\end{align}

We write this equation using regularising parameter $\beta$ as follows

\begin{align}
	=\lim_{\beta\rightarrow 1}\frac{1}{6^4}\prod_{v\in V}\Bigg(\Big(\beta\q^{(n^v_1+n^v_2)}-\q^{n^v_2}-\q^{n^v_1}+\q^{-n^v_1}+\q^{-n^v_2}-\q^{-(n^v_1+n^v_2)}\Big)^{2-\text{deg }v}\times\nonumber\\\Big(\q^{(n^v_1+n^v_2)}-\beta\q^{n^v_2}-\q^{n^v_1}+\q^{-n^v_1}+\q^{-n^v_2}-\q^{-(n^v_1+n^v_2)}\Big)^{2-\text{deg }v}\times\nonumber\\\Big(\q^{(n^v_1+n^v_2)}-\q^{n^v_2}-\beta\q^{n^v_1}+\q^{-n^v_1}+\q^{-n^v_2}-\q^{-(n^v_1+n^v_2)}\Big)^{2-\text{deg }v}\times\nonumber\\\Big(\q^{(n^v_1+n^v_2)}-\q^{n^v_2}-\q^{n^v_1}+\beta\q^{-n^v_1}+\q^{-n^v_2}-\q^{-(n^v_1+n^v_2)}\Big)^{2-\text{deg }v}\times\nonumber\\\Big(\q^{(n^v_1+n^v_2)}-\q^{n^v_2}-\q^{n^v_1}+\q^{-n^v_1}+\beta\q^{-n^v_2}-\q^{-(n^v_1+n^v_2)}\Big)^{2-\text{deg }v}\times\nonumber\\\Big(\q^{(n^v_1+n^v_2)}-\q^{n^v_2}-\q^{n^v_1}+\q^{-n^v_1}+\q^{-n^v_2}-\beta\q^{-(n^v_1+n^v_2)}\Big)^{2-\text{deg }v}\Bigg),
	\label{beta1}
\end{align}

now we expand the term inside the parenthesis in the above equation as $|\beta|<1$, to get the following:

\begin{equation}
	=\lim_{\beta\rightarrow 1}\left(\frac{1}{6^4}\sum_{s\in Q^4+\delta}\xi_s^\beta\q^{(\lambda,s)}\right)
\end{equation}

Above equation fixes the $\xi_s^1$. Moreover, $\delta_v=0~ \forall ~v$, for $\mathfrak{su}(3)$ Lie algebra as $\rho(=\Lambda_1+\Lambda_2=\alpha_1+\alpha_2)\in Q$. Further for this plumbing graph there is only one homological block which corresponds to $b=0$. Using all this we obtain the $\hat{Z}$-invariant (\ref{zhatdef}) as follows:

\begin{align}
	\hat{Z}_0^{\mathfrak{su}(3)}[M(\Gamma);q]=q^{3/2}\Big(1 - 2 q + 2 q^3 + q^4 - 2 q^5 - 2 q^8 + 4 q^9 + 2 q^{10} - 4 q^{11} + 2 q^{13} - 6 q^{14} + 2 q^{15} - 2 q^{16} + 4 q^{18} - q^{20}\nonumber\\ + 4 q^{21} - 2 q^{22} - 4 q^{23} + 2 q^{24} + 2 q^{25} - 4 q^{26} + 6 q^{30} - 2 q^{31} + 6 q^{33} - 2 q^{34} - 2 q^{35} - 2 q^{38} + q^{40} +\mathcal{O}(q^{41})\Big).
\end{align}

Note that the variable $q$ here is the analytically continued variable inside the unit circle.

\section{Sublattice $P'$ and Chern-Simons level $k'$ for $SU(4)/\mathbb{Z}_2$,$SU(6)/\mathbb{Z}_2$ and $SU(6)/\mathbb{Z}_3$}
\label{appendB}

In this appendix, we will present the sublattice $P'$ and Chern-Simons level $k'$ for some non-simply connected groups. The $i^{\text{th}}$ fundamental weight vector for $\mathfrak{su}(N)$ Lie algebra is given by:

\begin{equation}
\Lambda_i=\frac{1}{N}(\underbrace{N-i,N-i,\ldots,N-i}_{i-\text{times}},\underbrace{-i,-i,\ldots,-i}_{(N-i)-\text{times}})~~~\text{where}~i\in\{1,2,\ldots,N-1\}.	
\end{equation}

\paragraph{\underline{$SU(4)/\mathbb{Z}_2:$}}

The center of $SU(4)$ is $\mathbb{Z}_4$ which is isomorphic to $\{e,\Lambda_1,\Lambda_2,\Lambda_3\}$. Further, the center of $SU(4)/\mathbb{Z}_2$ is $\mathbb{Z}_2$ which is isomorphic to $\{e,\Lambda_2\}$. The root lattice $Q$ of $\mathfrak{su}(4)$ Lie algebra, which also corresponds to the equivalence class for identity group element, is given by adding the following vectors:
\begin{equation}
	\{(2n_1-n_2)\Lambda_1,(-n_1+2n_2-n_3)\Lambda_2,(-n_2+2n_3)\Lambda_3|n_1,n_2,n_3\in \mathbb{Z}\},
	\label{su4z2a}
\end{equation}
where $\Lambda_1,\Lambda_2$ and $\Lambda_3$ are weight vectors. The equivalence class for the group element $\Lambda_2$ corresponds to 
\begin{equation}
	\{(2n_1-n_2)\Lambda_1,(-n_1+2n_2-n_3+1)\Lambda_2,(-n_2+2n_3)\Lambda_3|n_1,n_2,n_3\in \mathbb{Z}\}.
	\label{su4z2b}
\end{equation}

Hence the lattice $P'$ is given by taking the union of two equivalence classes (\ref{su4z2a}) and (\ref{su4z2b}).

Chern-Simons level $k'=\gamma k+4$ and factor $\gamma$ is fixed by smallest integer which satisfies the following equation:

\begin{equation}
	\frac{\gamma}{2}(\Lambda_2,\Lambda_2)\in\mathbb{Z}\implies \gamma=2,
\end{equation}

therefore, $k'=2k+4$.

\paragraph{\underline{$SU(6)/\mathbb{Z}_2:$}}

The center of $SU(6)/\mathbb{Z}_2$ is $\mathbb{Z}_3\cong\{e,\Lambda_2,\Lambda_4\}$. Therefore the sublattice $P'$ corresponding to the allowed representations of $SU(6)/\mathbb{Z}_2$ is given by:
\begin{align}
	\{(2 n_1 - n_2) \Lambda_1 , (-n_1 + 2 n_2 - n_3) \Lambda_2 , (-n_2 + 2 n_3 - n_4) \Lambda_3 , (-n_3 + 2 n_4 - n_5) \Lambda_4 , (-n_4 + 2 n_5) \Lambda_5\}\cup\nonumber\\
	\{(2 n_1 - n_2) \Lambda_1 , (-n_1 + 2 n_2 - n_3+1) \Lambda_2 , (-n_2 + 2 n_3 - n_4) \Lambda_3 , (-n_3 + 2 n_4 - n_5) \Lambda_4 , (-n_4 + 2 n_5) \Lambda_5\}\cup\nonumber\\
	\{(2 n_1 - n_2) \Lambda_1 , (-n_1 + 2 n_2 - n_3) \Lambda_2 , (-n_2 + 2 n_3 - n_4) \Lambda_3 , (-n_3 + 2 n_4 - n_5+1) \Lambda_4 , (-n_4 + 2 n_5) \Lambda_5\}.
\end{align}

Chern-Simons level $k'=\gamma k+6$ where $\gamma$ is fixed by requiring:

\begin{equation}
	\frac{\gamma}{2}(\Lambda_3,\Lambda_3)\in\mathbb{Z}\implies 4,
\end{equation}

therefore, $k'=4k+6$.

\paragraph{\underline{$SU(6)/\mathbb{Z}_3:$}}

For $\mathbb{Z}_2\cong\{e,\Lambda_3\}$, $P'$ is:
\begin{align}
	\{(2 n_1 - n_2) \Lambda_1 , (-n_1 + 2 n_2 - n_3) \Lambda_2 , (-n_2 + 2 n_3 - n_4) \Lambda_3 , (-n_3 + 2 n_4 - n_5) \Lambda_4 , (-n_4 + 2 n_5) \Lambda_5\}\cup\nonumber\\
	\{(2 n_1 - n_2) \Lambda_1 , (-n_1 + 2 n_2 - n_3) \Lambda_2 , (-n_2 + 2 n_3 - n_4+1) \Lambda_3 , (-n_3 + 
	2 n_4 - n_5) \Lambda_4 , (-n_4 + 2 n_5) \Lambda_5\}.
\end{align}

and Chern-Simons level $k'=\gamma k+6$ is fixed as follows:

\begin{equation}
	\frac{\gamma}{2}(\Lambda_2,\Lambda_2)\in \mathbb{Z}\;\; \text{and}\;\; 	\frac{\gamma}{2}(\Lambda_4,\Lambda_4)\in \mathbb{Z} \implies \gamma=3,
\end{equation}

hence, $k'=3k+6$.

\section{GPPV conjecture for simply connected case: $SU(N)$}
\label{appenc}

In order to compare the GPPV conjecture with the simply connected $SU(N)$ group, we present the GPPV conjecture for the $SU(N)$ group. For the $SU(N)$ group, with the weight lattice being $P$ and the root lattice being $Q$, the WRT invariant for plumbed 3-manifold $M(\Gamma)$ can be decomposed as follows:

\begin{align}
	\tau^{SU(N)}_{k'}[M(\Gamma);\q]=\frac{1}{|W||\text{det} B|^{1/2}\sum_{w\in W}(-1)^{\ell(w)}\q^{(\rho,w(\rho))}}\sum_{a\in (Q)^L/B(Q)^L}\exp(-\pi
	ik'(a,B^{-1}a))\times\nonumber\\ \nonumber\\
	\sum_{b\in (Q^L+\delta)/BQ^L}\exp(-2\pi i(a,B^{-1}b))\lim_{q\rightarrow \q}\hat{Z}^{\mathfrak{su}(N)}_b[M(\Gamma),q].
\end{align}

where $B$ is the linking matrix for $\Gamma$. For a general closed 3-manifold with $b_1(M)=0$, we conjecture the following:

\begin{align}
	\tau^{SU(N)}_{k'}[M;\q]=\frac{1}{|H_1(M;\mathbb{Z})|^{\frac{N-1}{2}}\sum_{w\in W}(-1)^{\ell(w)}\q^{(\rho,\omega(\rho))}}\sum_{a,b\in (\text{Spin}^c(M))^{(N-1)}/S_N}\exp(-2\pi i k'\sum_{i=1}^{N-1}\ell k(a_i,a_i))\times\nonumber\\\nonumber\\
	\exp(-4\pi i\sum_{i=1}^{N-1}\ell k(a_i,b_i))\lim_{q\rightarrow\q}\hat{Z}_b^{\mathfrak{su}(N)}[M;q].
\end{align}

\end{appendices}

\bibliography{references.bib}

\end{document}